\documentclass{emulateapj}

\usepackage{amsmath}
\usepackage{gensymb}
\usepackage{color}

 \newcommand{\bra}[1]{\left( #1 \right)}
 \newcommand{\sqb}[1]{\left[ #1 \right]}

\DeclareMathOperator\erf{erf}

\shorttitle{Predicted Extension of the Sagittarius Stream to the Milky Way Virial Radius}
\shortauthors{M. Dierickx}

\begin{document}

\title{Predicted Extension of the Sagittarius Stream to the Milky Way Virial Radius}
\author{Marion I. P. Dierickx\altaffilmark{1} and Abraham Loeb\altaffilmark{1}}
\altaffiltext{1}{Astronomy Department, Harvard University, 60 Garden Street, Cambridge, MA 02138, USA; mdierickx@cfa.harvard.edu, aloeb@cfa.harvard.edu}

\begin{abstract}
The extensive span of the Sagittarius (Sgr) stream makes it a promising tool for studying the Milky Way gravitational potential. Characterizing its stellar kinematics can constrain halo properties and provide a benchmark for the Cold Dark Matter galaxy formation paradigm. Accurate models of the disruption dynamics of the Sgr progenitor are necessary to employ this tool. Using a combination of analytic modeling and N-body simulations, we build a new model of the Sgr orbit and resulting stellar stream. In contrast to previous models, we simulate the full infall trajectory of the Sgr progenitor from the time it first crossed the Milky Way virial radius 8 Gyr ago. An exploration of the parameter space of initial phase-space conditions yields tight constraints on the angular momentum of the Sgr progenitor. Our best-fit model is the first to reproduce accurately existing data on the 3D positions and radial velocities of the debris detected 100~kpc away in the MW halo. In addition to replicating the mapped stream, the simulation also predicts the existence of several arms of the Sgr stream extending to hundreds of kiloparsecs. The two most distant stars known in the Milky Way halo coincide with the predicted structure. Additional stars in the newly predicted arms can be found with future data from the {\it Large Synoptic Survey Telescope}. Detecting a statistical sample of stars in the most distant Sgr arms would provide an opportunity to constrain the Milky Way potential out to unprecedented Galactocentric radii.
\end{abstract}

%%%%%%%%%%%%%%%%%%%%%%%%%%%%%%
\section{Introduction}
\label{sec:intro}

The Local Group provides a natural laboratory for near-field cosmology and the study of galaxy formation. Dwarf satellites within the Local Group can be characterized at a high level of detail, providing a rich dataset to answer open questions about the structure of dark matter haloes. With a Galactocentric distance of only $\sim25$~kpc, the Sagittarius (Sgr) dwarf spheroidal is one of the nearest dwarf galaxies \citep{kunder09}. First discovered by \citet{ibata94}, Sgr is in a near polar orbit around the Milky Way (MW) and has experienced multiple passages through the disk \citep[e.g.][]{law10, purcell11}. The resulting stream of tidally stripped stars wraps a full 360$^{\degree}$ around the celestial sphere. Coincidentally, the Sun's location is close enough to the Sgr orbital plane to likely lie within the width of the debris trail \citep{majewski03}. 

Starting with \citet{johnston95}, many studies have attempted to constrain both the properties of Sgr and the MW based on the Sgr debris system. Building on increasingly detailed surveys of the stellar stream's leading and trailing arms, many investigations use the kinematics of the tidal debris as a diagnostic of the gravitational potential \citep[e.g.][]{ibata01, helmi04, law05, belokurov06, penarrubia06, law10, deg13, veraciro13, pricewhelan13,sohn15}. This approach has yielded ambiguous results \citep{ibata13}, alternatively pointing to prolate \citep{helmi04}, oblate \citep{johnston05}, spherical \citep{ibata01, fellhauer06},  or triaxial \citep{law09, law10, deg13} halo shapes.

Reconstructing the Sgr orbital history consistently with the observed stream necessarily underlies any inferences made about the MW's potential. However, as demonstrated by \citet{jiang00}, different families of orbital histories are allowed depending on the mass of the Sgr progenitor, due to dynamical friction. A more massive Sgr progenitor ($\sim 10^{11}$~M$_\odot$) would fall in from larger Galactocentric distances ($\gtrsim 200$~kpc) and undergo stronger mass stripping and dynamical friction. Conversely, the Sgr progenitor may have been as light as $\sim 10^{9}$~M$_\odot$ provided its initial separation was comparable to current apocentric distances of $\sim 60$~kpc. The uncertainty in the Sgr progenitor structure therefore goes hand in hand with the uncertainty in its orbital history. Properties of the Sgr dwarf's baryonic components, such as disk rotation, have also been shown to affect features of the resulting tidal stream \citep{penarrubia10}.

Recently, the two most widely referenced orbital models for Sgr have been those of \citet{law10} and \citet{purcell11}. The orbit introduced by \citet{purcell11} and used by \citet{gomez15} starts with the Sgr progenitor only 80~kpc away from the Galactic center, well within the virial radius of the MW. Earlier phases of the orbit are not simulated and the Sgr progenitor is artificially truncated at its instantaneous Jacobi radius in order to mimic tidal stripping during the early infall stage. However, the initial phase-space coordinates of the progenitor at 80~kpc are taken from Sgr example orbits by \citet{keselman09}, and not directly based on observable quantities. Therefore the resulting trajectory shares qualitative properties with the true Sgr orbit, but is unlikely to be an accurate match. 

The detailed work of \citet{law10} integrates the orbit of a test particle from Sgr's current location back in time in a fixed MW-like potential, under the assumption that Sgr is currently moving towards the galactic plane. This model does not utilize the proper motion from precision {\it Hubble Space Telescope} astrometry \citep{dinescu05,pryor10,massari13}. The proper motion predicted by the \citet{law10} model is within $~2\sigma$ of the estimates by  \citet{dinescu05} and \citet{pryor10}, but better accuracy could likely be achieved by basing the model directly on the transverse velocity measurements. Importantly, inferring the orbit from evolving a test particle backwards in time does not capture tidal stripping effects on the progenitor halo. This approach is valid in the low-mass, low dynamical friction, low stripping regime outlined by \citet{jiang00}, but cannot recover earlier infall phases for a progenitor with mass $\gtrsim 10^9$~M$_\odot$. The initial mass of Sgr used by \citet{law10} is $6.4 \times 10^8$~M$_\odot$, distinctly in the regime where dynamical friction is unimportant. 

Since these early efforts to model the Sgr orbit, there has been mounting evidence in favor of a more massive Sgr progenitor \citep{niederste10, conroy09, behroozi10, gibbons16}. This calls for a renewed effort to model the Sgr orbit, simultaneously accounting for higher progenitor mass and initial separation. In this paper, we set out to find a model for the full infall of the Sgr dwarf into the MW, until it reaches its current observed position and velocity. This includes the early infall phase at Galactocentric radii $> 60$-80~kpc not simulated by \citet{law10}, \citet{purcell11} or \citet{gomez15}. The survival of the Sgr satellite until the present day implies that it cannot have formed deep inside the MW halo, where it would have been cannibalized already. Rather, hierarchical models of galaxy formation suggest that the Sgr dwarf likely formed early on in the periphery of the assembling host halo. We therefore aim to initiate the Sgr progenitor at the MW's virial radius at redshift of $z = 1$, approximately 8~Gyr ago. The higher initial separation and Sgr mass preclude integrating orbits backwards in time and call for full forward modeling, including tidal stripping and dynamical friction. We therefore use a combination of analytic and N-body modeling in a dual approach to test possible trajectories for Sgr. The high computational cost of full N-body simulations, where both the MW and Sgr progenitor are modeled with live haloes, prohibits a brute force exploration of the parameter space. Therefore, we first carry out a systematic search of the parameter space with a fast semi-analytic, point particle-like model in Section \ref{sec:SAM}. We then perform N-body simulations of the best-fit models with {\footnotesize GADGET} for comparison with the Sgr tidal stream in Section \ref{sec:nbody}. Finally, we discuss our main conclusions in Section \ref{sec:conclusions}.

%%%%%%%%%%%%%%%%%%%%%%%%%%%%%%
\section{Semi-Analytic Model}
\label{sec:SAM}

\subsection{Galaxy Parameters}
\label{subsec:samgalaxyparams}

\begin{deluxetable}{llll}
\tablewidth{0pt}
\tablecaption{Galaxy Parameters for Semi-Analytic and N-Body Simulations.}
\tablehead{Parameter & Description & MW & Sgr dSph}
\startdata
$M_{\rm NFW}$ & NFW halo mass & $1\times10^{12}$ M$_{\odot}$& $1\times10^{10}$ M$_{\odot}$ \\
$c$ & NFW halo concentration & 10 & 8 \\
$M_{\rm halo}$ & Hernquist total mass & $1.25\times10^{12}$ M$_{\odot}$& $1.3\times10^{10}$ M$_{\odot}$ \\
& Particles in halo & $1.16 \times10^6$ & $1.17 \times10^4$ \\
$R_{200c, z=0}$ & Present-day virial radius & 206 kpc & 44 kpc \\
$r_\text{H}$ & Hernquist scale radius & 38.35 kpc & 9.81 kpc \\
$M_{\rm disk}$ & Disk mass & 0.065 $M_{\rm halo}$  & 0.06 $M_{\rm halo}$   \\
& Particles in disk & $2.03 \times10^6$ & $1.95 \times10^4$ \\
$M_{\rm bulge}$ & Bulge mass & 0.01 $M_{\rm halo}$ & 0.04 $M_{\rm halo}$  \\
& Particles in bulge & $3.125 \times10^5$ & $1.3 \times10^4$ \\
$b_{\rm 0}$ & Disk scale length & 3.5 kpc & 0.85 kpc \\
$c_{\rm 0}$ & Disk scale height & 0.15 $b_{\rm 0}$ & 0.15 $b_{\rm 0}$ \\
$a$ & Bulge scale length & 0.2 $b_{\rm 0}$ & 0.2 $b_{\rm 0}$ \\
\hline \hline
\enddata
\tablecomments{MW parameters are adapted from \citet{gomez15}. The parameters for Sgr are determined based on several considerations: the virial mass is chosen as estimated by \citet{niederste10}, the disk scale radius following \citet{gomez15}, and a stellar mass percentage larger than in \citet{gomez15} by a factor of 3 for practical (resolution) reasons. We use standard $\Lambda$CDM cosmological parameters values of $H_{0} = 70$~km~s$^{-1}$~Mpc$^{-1}$ and $\Omega_{\rm m} = 0.27$ \citep{planck15}.
\label{tab:params}}
\end{deluxetable}

Galaxy formation is a continuous process and as a result, the formation redshift of galaxies such as the MW and Sgr dwarf cannot be pinpointed exactly. We aim to trace the system's history out to a redshift of $z = 1$, corresponding to approximately 8~Gyr of lookback time. This choice is based on the age of the M-giants in the stream, estimated by \citet{bellazzini06} to be $8.0\pm1.5$~Gyr. We adopt this evolutionary timescale as the guiding principle for our initial conditions, postulating that the Sgr progenitor may first have crossed the MW's virial radius around that time. We adopt a fiducial virial mass of $10^{12}$~M$_\odot$ for the MW, consistent with recent estimates \citep[e.g.][]{xue08}. Various simulations of halo assembly histories suggest that today's MW-like haloes would have built up at least half of their mass by $z = 1$, with a significant scatter of $\sim 2\times10^{11}$~M$_\odot$ depending on the halo's accretion history \citep[e.g.][]{torrey15, lu16}.  Using the spherically symmetric top-hat framework of halo formation, the virial radius of a $5\times10^{11}$~M$_\odot$ galaxy collapsing at $z = 1$ is approximately 124~kpc \citep{barkana01}. We therefore initiate the Sgr progenitor at a Galactocentric radius of 125~kpc and aim to trace its evolution over a period of 8~Gyr. 

Halo growth is believed to occur inside-out, with later additions of mass being appended on the outskirts of the halo \citep[e.g.][]{loeb03}. This picture has been validated in simulations \citep[e.g.][]{wellons16} and observations \citep[e.g.][]{delarosa16} of high-redshift massive compact galaxies. Cosmological simulations also show that after $z=1$, most MW analogs do not undergo a major merger \citep[e.g.][]{fakhouri10} and have inner halo profiles that remain essentially fixed \citep[e.g.][]{gao04, mollitor15}. Studies of the stellar populations in the MW disk suggest a quiet evolution since $z = 2$, with no significant mergers in the last $\sim10$~Gyr \citep{wyse01,hammer07}. The Sgr orbit in our model is contained inside Galactocentric radii $<125$~kpc and is consequently not impacted by the later addition of matter outside of this sphere (assuming spherical symmetry). As a result, we maintain the same initial halo profile throughout each simulation. A halo's scale radius $r_s$ is related to its virial radius $R_\text{vir}$ and concentration parameter $c$ by the relationship $r_s = R_\text{vir}/c$. As the halo accretes mass on its outskirts, the scale radius is maintained constant while the virial radius and the concentration parameter grow. Following the concentration growth of galaxies in the models of \citet{diemer15}, a MW analog would have grown from $c \simeq 7$, $R_\text{vir} \simeq 125$~kpc at $z = 1$ to $c \simeq 10$, $R_\text{vir} \simeq 200$~kpc at $z = 0$, giving a  nearly constant scale radius of $\sim$18-20~kpc. For a $z = 0$ MW analog (with parameters as outlined in Table~\ref{tab:params}), the mass inside a radius of 125~kpc is approximately $7\times10^{11}$~M$_\odot$, consistent with the virial mass of $(5\pm 2)\times10^{11}$~M$_\odot$ of its $z=1$ progenitor. Beyond the prescriptions for the MW's evolution adopted here, \citet{penarrubia06} provide an in-depth study of modeling tidal streams in evolving dark matter halos. Their simulations suggest that the debris configuration reflects the present galactic potential shape rather than its evolution. Our choice of a constant host potential profile throughout the orbit of Sgr is therefore unlikely to have a strong effect on the phase-space distribution of the stream.

% Figure: Hernquist / NFW mass matching
\begin{figure*}[hbt]
\begin{center}
\includegraphics[scale=0.4]{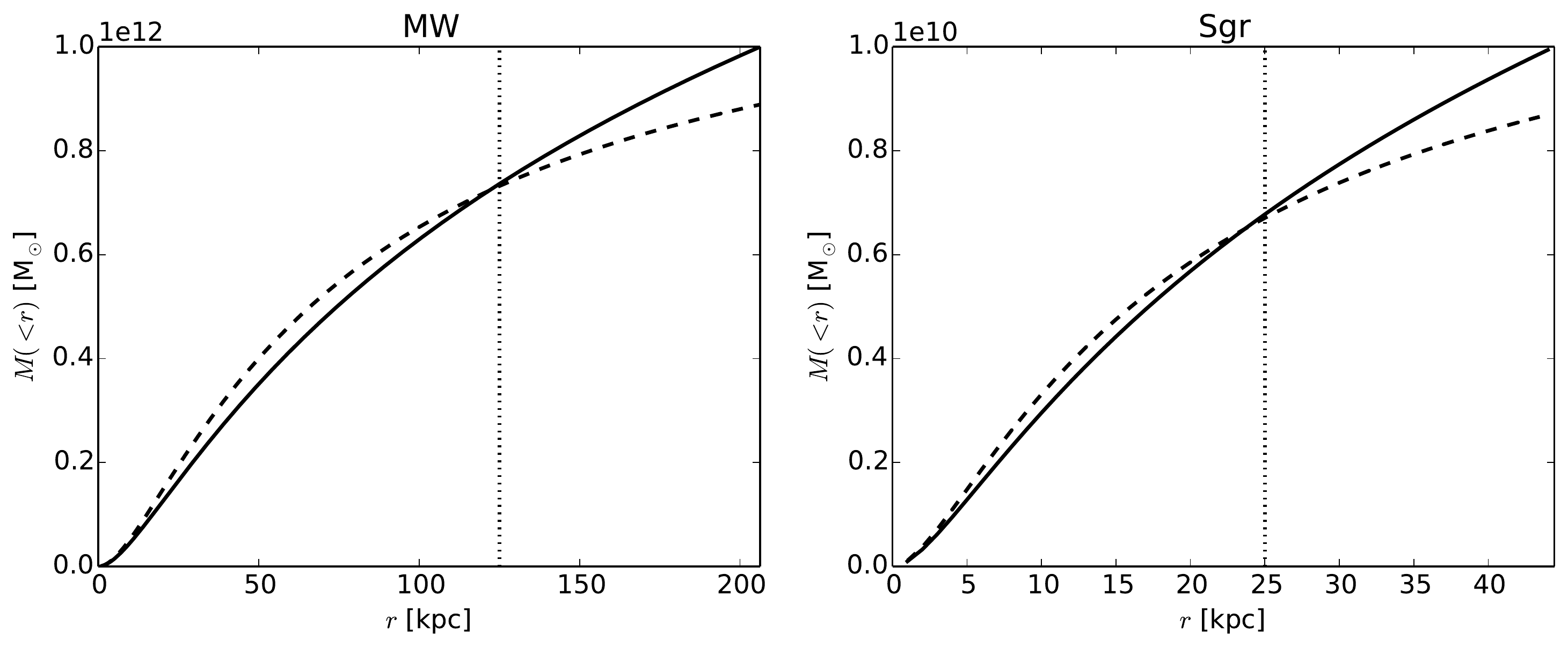} 
\caption{Enclosed mass profiles for the chosen halo profiles of the MW and Sgr. The solid line shows the reference NFW profile \citep{navarro97} for each galaxy. Dotted lines indicate the boundary radii where the enclosed mass values are matched: $d_{\text{init}} = 125$~kpc for the MW, and $r_{\text{tidal, init}} \simeq 25$~kpc for Sgr. The dashed lines show the resulting Hernquist approximations to the NFW profiles. Profile parameters are provided in Table~\ref{tab:params}.
\label{fig:massmatching} }
\end{center}
\end{figure*}

The MW potential includes three components: a dark matter halo and bulge following Hernquist profiles \citep{hernquist90}, and an exponential disk:
\begin{equation}
\Phi_{\text{MW}} = \Phi_{\text{halo}} + \Phi_{\text{disk}} + \Phi_{\text{bulge}} \ ,
\end{equation}
where 
\begin{eqnarray}
& \Phi_{\text{halo}}(r) & = - \frac{G M_\text{halo}}{r+r_\text{H}}  \ , \\
& \Phi_{\text{disk}}(r) & = - G M_\text{disk} \bra{ \frac{1-e^{-r/b_\text{0}}}{r}}  \ , \\
& \Phi_{\text{bulge}}(r) & = - \frac{G M_\text{bulge}}{r+c_\text{0}} \ .
\end{eqnarray}
For simplicity, the disk and bulge components are omitted for Sgr in the semi-analytic model, but included in the full simulation described in Section~\ref{sec:nbody}. The parameters of both galactic potentials are summarized in Table~\ref{tab:params}. The parameters of the Hernquist halo potentials are chosen such that the mass enclosed within the radius of interest matches that of fiducial Navarro, Frenk \& White profiles \citep[NFW;][]{navarro97} profiles. The resulting enclosed mass curves are shown in Figure~\ref{fig:massmatching}. For the MW, we adjust the enclosed mass inside the initial distance between the galaxy centers, $d_{\text{init}}$ = 125~kpc. A total halo mass of $M_\text{halo} = 1.25 \times 10^{12}$~M$_\odot$ and a scale radius of $r_\text{H} = 38.35$~kpc matches the fiducial NFW halo with a virial mass $M_{\text{MW}} = 10^{12}$~M$_\odot$ and a concentration parameter of 10. For Sgr, we use $M_\text{halo} = 1.3 \times 10^{10}$~M$_\odot$ and $r_\text{H} = 9.81$~kpc. This corresponds to the mass enclosed within the initial tidal radius, $r_{\text{tidal, init}} \simeq 25$~kpc, of an NFW potential with $M_{\text{Sgr}} = 10^{10}$~M$_\odot$ and $c_{\text{Sgr}}$ = 8. 

Our choice of initial parameters for Sgr is based on the study by \citet{niederste10}, who reconstruct the properties of the progenitor by conducting a census of the stellar tidal debris. A lower bound of $(9.6-13.2)\times 10^7$~L$_\odot$ is inferred for the progenitor's luminosity by summing up the luminosities of the Sgr core, leading and trailing arms. Relating this value to results from cosmological $N$-body simulations, \citet{niederste10} estimate a mass of $\sim10^{10}$~M$_\odot$ for the Sgr dark matter halo prior to tidal disruption. Based on this choice of progenitor mass, the mass-concentration relation models of \citet{diemer15} suggest a low concentration of $c\simeq8$ at $z = 1$. We then use these parameters as our reference NFW profile, and tune the Hernquist profile parameters to match the mass enclosed inside the initial tidal radius of the system (see Figure~\ref{fig:massmatching}). The resulting Hernquist scale radius $r_\text{H} = 9.81$~kpc is consistent with the values used by \citet{gomez15} for their 'Light Sgr' model, which has a mass of $3.2\times10^{10}$~M$_\odot$. Since only one wrap of the debris is considered and additional luminous matter may be located at apocenter pile-ups, the true Sgr progenitor mass may exceed the lower bound established by \citet{niederste10}. Given a total luminosity of $\sim 10^8$~L$_\odot$, cosmological abundance matching would suggest a higher mass of $\sim10^{10.5}-10^{11}$~M$_\odot$ \citep{conroy09, behroozi10}. Exploring a range of different progenitor models, \citet{gibbons16} find that masses $\gtrsim6 \times 10^{10}$~M$_\odot$ are most consistent with the velocity dispersion of the stream stellar populations. Still, the orbit presented here is the first physically motivated model for the Sgr infall that includes a progenitor mass closer to the most recent mass estimates. 

\subsection{Methodology}
\label{subsec:sammethods}

% Figure: Initial setup
\begin{figure*}[hbt]
\begin{center}
\includegraphics[scale=0.5]{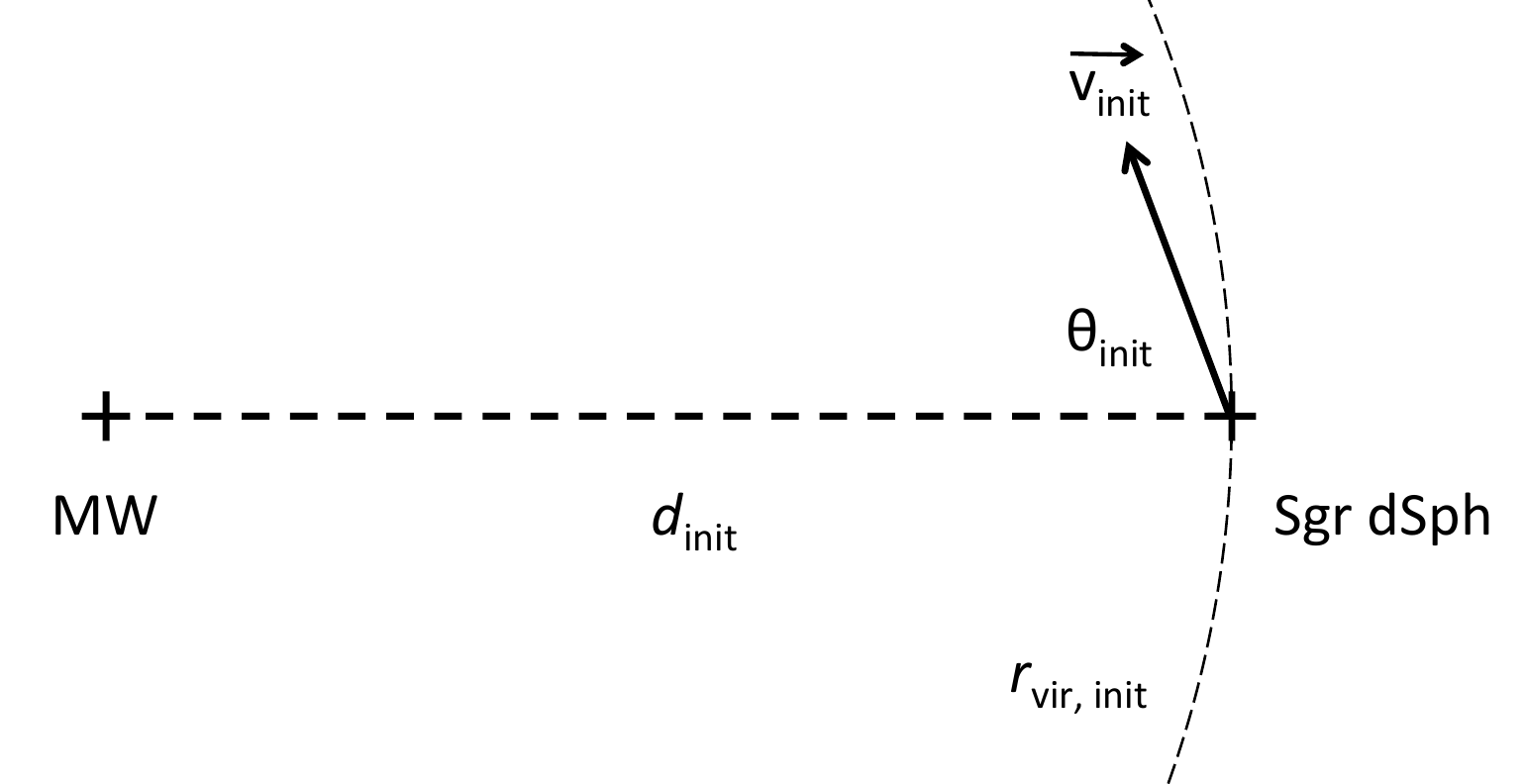} 
\caption{The initial configuration of the MW and Sgr dSph progenitor in the semi-analytic orbital model. Sgr is initially located a distance $d_{\text{init}}$ away from the center of the Galaxy, corresponding to the virial radius of the MW at that time ($z \sim 1$). Sgr is given an initial velocity vector $\vec{v}_{\text{init}}$ with variable magnitude and angle $\theta_{\text{init}}$.
\label{fig:setup} }
\end{center}
\end{figure*}

In our search for an orbital model for the Sgr dwarf spheroidal, we start testing possible trajectories with an approximate point-particle approach. From predetermined initial conditions, the equations of motion are solved numerically forward in time with a fourth order Runge-Kutta method. Both the dynamics of Sgr in the MW potential and the MW's response to the gravitational attraction of Sgr are modeled, including tidal stripping and dynamical friction for Sgr. We emphasize the fact that the MW potential is not held fixed at the origin; rather, we account for the mutual gravitational attraction between the MW and Sgr and allow the MW to move freely about the common center of mass of the system. As the studies of \citet{dierickx14} and \citet{gomez15} have shown, the response of a MW-like host galaxy to an infalling satellite can be significant. This is particularly relevant in the high separation, high Sgr progenitor mass regime explored here.

Keeping the initial separation fixed, the remaining free parameters quantify the amount of initial orbital angular momentum we grant Sgr at its starting location. We specify two quantities: $v_{\text{init}}$, the magnitude of the Sgr velocity, and $\theta_{\text{init}}$, the angle between the velocity vector and the direction to the MW center (see Figure \ref{fig:setup}). We integrate a two-dimensional grid of 2500 orbits with $v_{\text{init}}$ ranging from 0 to $\sim320$~km~s$^{-1}$, the NFW escape velocity from the MW halo at the initial radius of the Sgr progenitor, and $\theta_{\text{init}}$ ranging from 10 to 90 degrees. With total energy per unit mass defined as $E = \frac{1}{2}v_\text{init}^2 - \Phi(d_\text{init})$, the upper limit on velocity investigated here corresponds to $E = 0$, i.e. a marginally bound Sgr falling into the MW from a larger distance. We specify a nonzero lower limit on the initial angle in order to avoid purely radial orbits, which cause pathological behavior in the model. Angles $>90^\circ$ are excluded as they would imply a Sgr progenitor velocity directed away from the MW. 

Throughout the orbital integration, the MW experiences acceleration from a standard Hernquist gravitational potential for a tidally-truncated Sgr. For the Sgr satellite however, the following acceleration is calculated at Galactocentric coordinates $\vec{r}$:
\begin{equation}
\vec{a} = - \frac{G M_{\text{halo}}(< r)}{r^3}\vec{r} -\nabla \Phi_{\text{disk}} - \nabla \Phi_{\text{bulge}} + \vec{a}_{\text{DF}} \ ,
\end{equation}
where $M_{\text{halo}}(< r)$ is the mass interior to radius $r$ of the host halo, given simply by $M_{\text{halo}}(< r) = M_{\text{halo}} [r^2 / (r+r_H)^2]$ \citep{hernquist90}. The acceleration introduced by dynamical friction, $\vec{a}_{\text{DF}}$, is modeled using the Chandrasekhar formula \citep[][eq. 7.18]{binney87}:
\begin{multline}
\vec{a}_{\text{DF}} = - \frac{4 \pi \ln(\Lambda) G^2 \rho(r) M_{\text{Sgr}}(< r_t) }{v^3} \times \\
\sqb{\erf(X) - \frac{2X}{\sqrt{\pi}} \exp(-X^2) } \vec{v}
\end{multline}
%\begin{equation}
%\vec{a}_{\text{DF}} = - \frac{4 \pi \ln(\Lambda) G^2 \rho(r) M_{\text{Sgr}}(< r_t) }{v^3} \times \sqb{\erf(X) - \frac{2X}{\sqrt{\pi}} \exp(-X^2) } \vec{v} \ ,
%\end{equation}
where $\rho(r)$ is the density of the host halo at Galactocentric distance $r$, $ M_{\text{Sgr}}(< r_t)$ is the Sgr mass interior to its minimum tidal radius $r_t$, and $\ln(\Lambda)$ is the Coulomb logarithm. The remaining term involves $X = v/\sqrt{2}\sigma$, where $v$ is the satellite velocity and $\sigma$ is the one-dimensional velocity dispersion of particles in the host halo \citep[given by][eq. 10]{hernquist90}. Based on the formalism used by \citet{besla07} in their study of the orbital evolution of the Magellanic Clouds, we adopt an alternative time-dependent Coulomb logarithm as follows \citep{hashimoto03}:
\begin{equation}
\ln (\Lambda) = \ln \bra{\frac{r}{1.4 \epsilon}} \ ,
\end{equation}
where $\epsilon$ is a softening length variable used by \citet{hashimoto03} to model the Large Magellanic Cloud (LMC) with a Plummer sphere. This time-dependent parameterization of the Coulomb logarithm is also in agreement with the calibration carried out by \citet{vandermarel12b} for the M31-M33 interaction (which is of similarly unequal mass). We ran a series of semi-analytic orbits to test the friction parameterization, and find that setting $\epsilon \simeq 1$~kpc gives the best agreement with $N$-body integrations of the same initial conditions. This halo parameter is smaller by a factor of 3 compared to the parameterization used by \citet{hashimoto03} for the LMC, not surprising given that the mass ratio between host and satellite is more unequal by a factor of 10 in our case. Comparing to the standard formalism where $\Lambda = b_\text{max}/b_\text{min}$ \citep{binney87}, we have in essence set the impact parameter $b_\text{min}$ to $\sim1.4$~kpc (a reasonable value given the large range of initial conditions explored here) and introduced a time dependence for the cutoff radius $b_\text{max}$. These adjustments are generally expected to have a minor effect given their logarithmic contribution to the dynamical friction exerted on Sgr.

Sgr experiences tidal forces as it moves across the MW halo, leading to the outermost layers of material being stripped. The instantaneous tidal radius for Sgr, $r_t$, is found by solving the following equation numerically \citep{king62}:
\begin{equation}
r_t = r \sqb{\frac{1}{2} \frac{M_{\text{Sgr}}(< r_t)}{M_{\text{MW}} (< r)} }^{1/3} \ .
\end{equation}
Since material that has been stripped is considered lost, the tidal radius is computed at each time step and if necessary, updated to the lowest value calculated so far. Using this prescription we model dynamical friction on a progressively less massive Sgr halo. Since the lost Sgr mass is negligible compared to the MW mass, we ignore its contribution to the MW potential.

From its initial location at $d_\text{init} = 125$~kpc with varying velocity vector $\vec{v}_\text{init}$, the orbit of the Sgr progenitor is integrated forward in time for 10~Gyr with 10~Myr time steps. The current position of Sgr can be described by its galactic longitude and latitude $(l,b) = (5.6^\circ, -14.2^\circ)$ \citep{majewski03} and its heliocentric distance $d_\text{helio} = 25\pm2$~kpc \citep{kunder09}. We define a right-handed Cartesian coordinate system centered on the Galactic Center, with the $z$-axis pointing towards the North Galactic Pole and with the Sun's current location at [-8, 0, 0]~kpc \citep{honma12, reid14}. The coordinates of Sgr in this system can then be found with the following simple transformations (in units of kiloparsecs):
\begin{eqnarray}
x &=& d_\text{helio} \cos b \cos l - 8 \ , \\
y &=& d_\text{helio} \cos b \sin l \ , \\
z &=& d \sin b \ .
\end{eqnarray}
Applying these equations, the current position vector of the center of Sgr is $\vec{r}_\text{Sgr, obs} = (16.1, 2.35, -6.12)$ kpc, in agreement with the values provided e.g. by \citet{law10}. The heliocentric radial velocity of Sgr has been measured at $140\pm0.33$~km~s$^{-1}$ \citep[weighted mean of the Sgr,N and M54 average velocities estimates in Table 5 of][]{bellazzini08}, and its proper motion in the equatorial coordinate system is $(\mu_\alpha, \mu_\delta) = (-2.95 \pm 0.18, -1.19 \pm 0.16)$~mas~yr$^{-1}$ \citep{massari13}. Using these heliocentric velocity components, the Cartesian, right-handed galactic space velocity $(U, V, W)$ is calculated following \citet{johnson87}. Finally, these heliocentric space velocities are converted to the Galactic Rest Frame (GSR) by adding contributions from the local standard of rest and solar peculiar motions \citep{schoenrich10, reid14}: 
\begin{equation}
\vec{v}_\text{GSR} [\text{km s}^{-1}] = (U, V, W) + (0, 237, 0) + (11.1, 12.24, 7.25) \ ,
\end{equation}
This yields a current GSR velocity vector for Sgr of $\vec{v}_\text{Sgr, obs} = (242.5, 5.6, 228.1)$~km~s$^{-1}$ and a total velocity magnitude of $333\pm30$~km~s$^{-1}$.

At every time step along the trajectory, the quality of the match to the present-day configuration is investigated. This is a first done in a spherically symmetric sense before finding a best-match orientation for the Sun's location. At every step, the MW-Sgr distance and the Sgr velocity magnitude in the galaxy rest frame are compared to the observed values outlined above. For the times when Sgr is within $3\sigma$ of the correct galactocentric distance and the correct velocity magnitude, the analysis progresses to the next step. At this stage, it is necessary to orient the system in order to further quantify the match to observables. We utilize the spherical symmetry of the host potential to identify which point on a 8~kpc-radius sphere around the Galactic Center provides the best match to the Sun's location. 

To this end, we construct a Fibonacci lattice of 1,600 points with radius 8~kpc centered at the Galactic Center. The Fibonacci lattice is a convenient method to generate any number $N$ of evenly distributed points on the surface of a sphere, with each point representing almost the same area \citep{gonzalez10}. Points are arranged along a tightly wound generative spiral, and using the golden angle ($\sim 137.5^\circ$) as the value for the longitudinal turn between consecutive points maximizes packing. With $N=$~1,600, every lattice point occupies on average $\sim25\deg^2$, representing a one-dimensional positional uncertainty of approximately $5\deg$. Scaled by the ratio of the Sgr heliocentric distance to the Sun's galactocentric distance (approximately a factor of 3), a lattice with $N =$1,600 points yields an angular uncertainty of $\sim1.5-2\deg$ in the position of Sgr in the sky without becoming computationally prohibitive. This is acceptable given the many approximations and the uncertainties associated with the parameters of the galactic potentials. 

% Table: Chi-squared parameters
\begin{table*}[bt]
\caption{Observable and Simulated Parameters Used in Chi-Squared Analysis.}
\begin{tabular}{lcccc}
\hline
Quantity & Observed Value & Semi-Analytic Model & $N$-body Model & Observational Error ($\sigma$) \\
\hline
(i) [kpc] & 17.4 & 18.6 & 21.0 &  2.0 \\
(ii) [kpc] & 25.0 & 26.0 & 27.6 & 2.0 \\
(iii) [$^\circ$] & 6.92 & 7.6 & 10.9 & 5 \\
(iv) [km~s$^{-1}$] & 178.8 & 180.8 & 172.8 & 1.5 \\
(v) [km~s$^{-1}$] & 281.0 & 353.8 & 285.0 & 30 \\
(vi) [$^\circ$] & 162.0 & 160.4 & 162.8 & 1.5 \\
\hline
Resulting $\chi^2_r$ & - & 1.32 & 2.28 & - \\
\hline
\hline
\end{tabular}
\tablecomments{The six reference quantities used in the $\chi^2$ analysis are provided along with the simulated parameters for both the best-match analytic and $N$-body models (see Section~\ref{sec:nbody}). The table rows are as follows: (i) GC - Sgr distance [kpc]; (ii) Sgr - Sun distance [kpc]; (iii) GC - Sgr - Sun angle [$^\circ$]; (iv) Sgr heliocentric line of sight velocity [km~s$^{-1}$]; (v) Sgr heliocentric transverse velocity [km~s$^{-1}$]; (vi) Angle between Sgr transverse velocity vector and the direction to the GC [$^\circ$].
\label{tab:chisquared}}
\end{table*}

For every candidate Sun position on the lattice, we compute the following 6 quantities: i) The Galactic Center - Sgr distance; ii) Sgr heliocentric distance; iii) the Galactic Center - Sgr - Sun angle; iv) the Sgr heliocentric radial velocity; v) the Sgr heliocentric transverse velocity magnitude; vi) the angle between the Sgr transverse velocity vector and the direction to the Galactic Center. The match to the corresponding measured numbers is quantified via a chi-squared test. The parameter values and associated errors used in the analysis are provided in Table~\ref{tab:chisquared}. The lattice point with the lowest reduced chi-squared is identified and recorded as the best-match position for the Sun at each favorable time step in the run. The moment along each trajectory with the lowest reduced chi-squared between 7 and 8.5~Gyr after initialization is extracted in order to compare different sets of initial conditions.

\subsection{Results of Semi-Analytic Modeling}
\label{subsec:samresults}

% Figure: Ridge at 10^10 M_sun figure
\begin{figure*}[hbt]
\begin{center}
\includegraphics[scale=0.6]{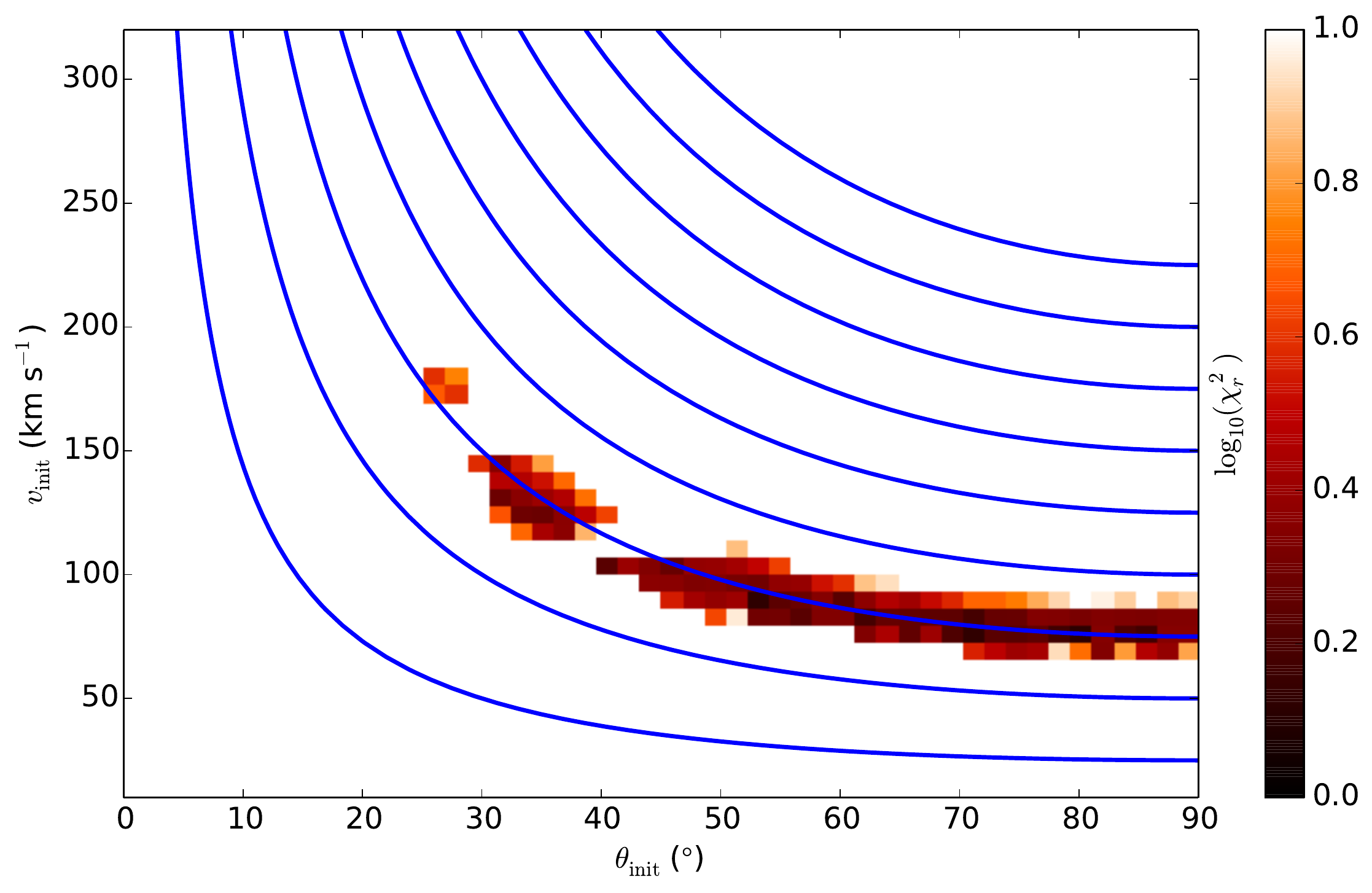} 
\caption{Quality of match to the Sgr observed phase-space coordinates over the initial parameter space considered in our semi-analytic model. The angle of the initial velocity vector away from the MW ($\theta_\text{init}$) is shown on the $x$-axis and the initial velocity magnitude ($v_\text{init}$) is plotted on the $y$-axis. Values of $\log_{10}(\chi^2_r) > 1$ are too large to be of interest and are uniformly colored white. Blue lines show contours of constant initial specific angular momentum per unit distance. (These are simply given by $v_\text{init} \sin\theta_\text{init} = \text{constant}$ since the initial MW-Sgr separation and Sgr progenitor mass are kept constant.) The contours range from 25 to 250~km~s$^{-1}$ for $\theta_\text{init} =90^\circ$ and are evenly spaced every 25~km~s$^{-1}$. 
\label{fig:ridge} }
\end{center}
\end{figure*}

Figure~\ref{fig:ridge} shows the reduced chi-squared values associated with the best-match snapshots along each trajectory, over the initial parameter space outlined in Section~\ref{subsec:sammethods}. In this color map, darker pixels indicate a closer match to the observed position and velocity of Sgr today. No matches are found for initial angles $\theta_\text{init}\lesssim 30^\circ$, indicating that nearly radial orbits are incompatible with the distance and velocity constraints imposed by the available data for Sgr. At high initial incidence angles ($\theta_\text{init}\gtrsim 60^\circ$), a specific range of initial velocities ($80$~km~s$^{-1} \lesssim v_\text{init} \lesssim 100$~km~s$^{-1}$) is preferred independently of the angle. Tangential velocities in this range are reasonable given analogous measurements of Local Group satellites: the M33 tangential velocity with respect to M31 is $\sim$129~km~s$^{-1}$  \citep{vandermarel12a}; the tangential velocities of the LMC and SMC with respect to the MW are $\sim$314 and $\sim$61~km~s$^{-1}$, respectively \citep{kallivayalil13}.

Contours of constant initial angular momentum are superimposed on the color map in Figure~\ref{fig:ridge}. Values of reduced chi-squared indicating a good match $(\log_{10}(\chi^2_r) \lesssim 0.5)$ are tightly grouped together in a ``valley'' of constant angular momentum. This suggests that in order to reach the current observed position and velocity of Sgr, a narrow range of initial angular momenta is preferred. The $v_\text{init}$ width of this range per unit mass and distance is approximately 20~km~s$^{-1}$. We thus conclude that the model favors a specific value for the angular momentum Sgr had 7-8~Gyr ago, upon crossing the MW virial radius for the first time. 

% Figure: Best-fit run from SAM
\begin{figure*}[hbt]
\begin{center}
\includegraphics[scale=0.55]{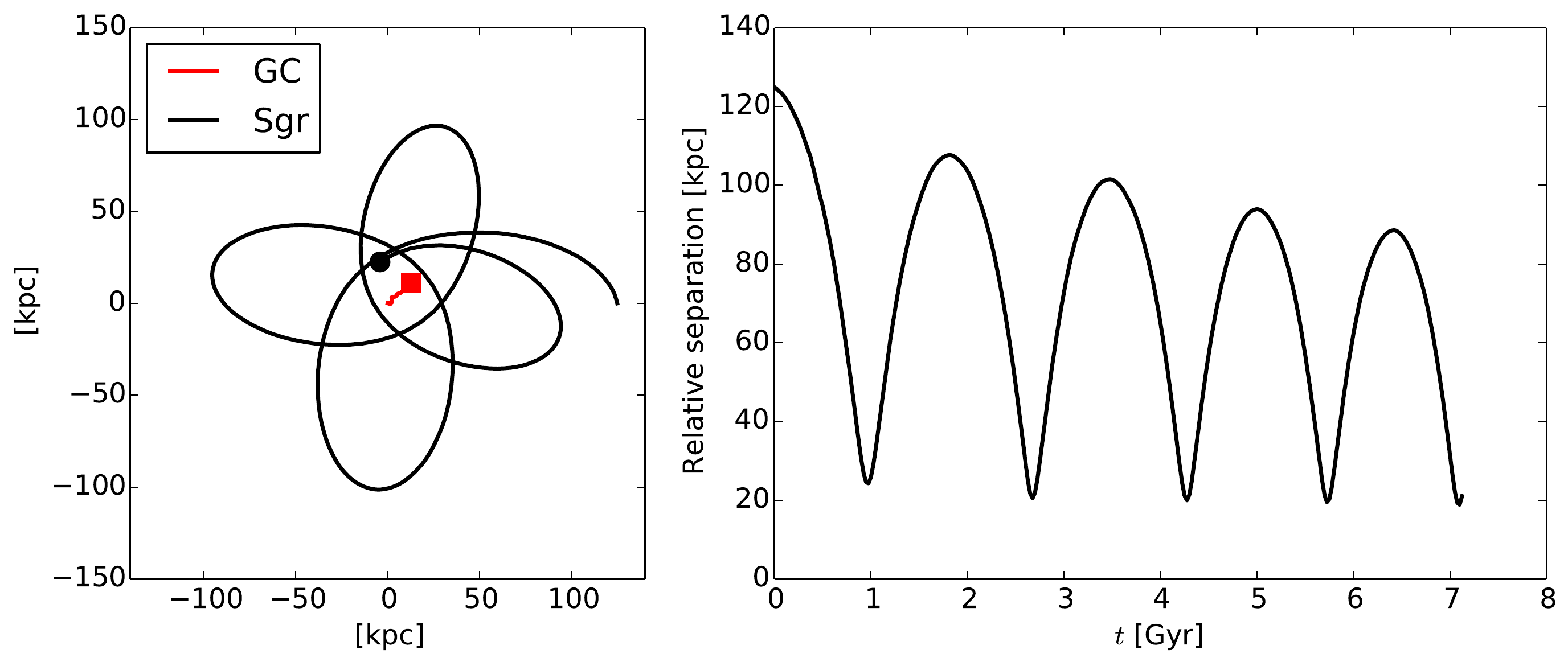} 
\caption{Overview of the best-fit Sgr orbit computed by the semi-analytic model described in Section~\ref{sec:SAM}. {\bf Left panel:} trajectories of the Sgr progenitor (red line) and MW Galactic Center (GC; black line) in the Sgr orbital plane. The current locations of the two galaxies are indicated by a colored dot and square, respectively. {\bf Right panel:} separation between the centers of the two galaxies (solid line) and the minimum tidal radius (dashed line) of Sgr as a function of time since the beginning of the calculation. The closest match to the current position and velocity of Sgr is reached after 7.71~Gyr. 
\label{fig:SAMtraj} }
\end{center}
\end{figure*}

% Figure: Best-fit run from SAM 3D arrow match
\begin{figure*}[hbt]
\begin{center}
\includegraphics[scale=0.5]{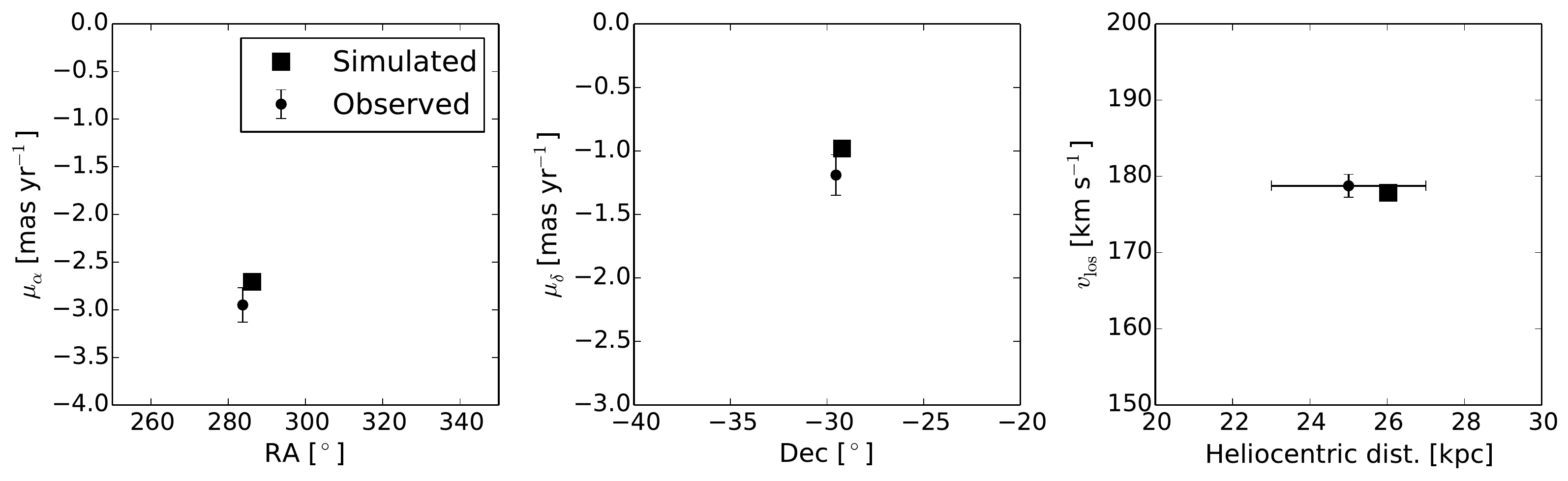} 
\caption{Comparison of the observed and simulated heliocentric coordinates of Sgr in the best-fit orbit from the semi-analytic model. The position and velocity are shown as projected on the sky in equatorial coordinates and along the line of sight. Error bars show the $1\sigma$ uncertainties associated with each measurement. The distance estimate is taken from \citet{kunder09}, the proper motion measurements from \citet{massari13}, and the line of sight velocity from \citet{bellazzini08}.
\label{fig:SAM3Dmatch} }
\end{center}
\end{figure*}

Next we extract the set of initial parameters with the lowest reduced chi-squared from Figure~\ref{fig:ridge}, and the resulting orbit is shown in Figure~\ref{fig:SAMtraj}. In terms of the parameters defined in Figure~\ref{fig:setup}, an initial velocity of $v_{\rm init}=72.6$~km~s$^{-1}$ directed at an angle of $\theta_{\rm init} = 80.8^\circ$ yields a $\chi^2_r$ value of 1.32 at a time 7.71~Gyr after the start of the integration (see Table~\ref{tab:chisquared}). The left panel of Figure~\ref{fig:SAMtraj} shows the resulting trajectory of the Sgr progenitor (in blue) and the MW's drift (in black). The right panel shows the separation between the two galaxies as a function of time (dashed line) as well as the minimum tidal radius of the Sgr galaxy calculated as described in Section~\ref{sec:SAM} (black line). Figure~\ref{fig:SAM3Dmatch} presents a comparison of the best-match simulated and measured phase-space coordinates of Sgr. Overall the semi-analytic model reproduces the position and velocity vector of Sgr very well.

%%%%%%%%%%%%%%%%%%%%%%%%%%%%%%
\section{N-Body Simulation}
\label{sec:nbody}
 
So far we have only used the phase-space coordinates measured for the Sgr remnant in order to constrain its infall history into the MW. However, additional data available for the Sgr stellar stream from various surveys provide an important way to assess how our model compares to observations. The semi-analytic model described in \S~\ref{subsec:sammethods}, while useful in exploring parameter space, does not generate mock debris streams. Importantly, we also wish to improve on the spherical symmetry approximation made previously by including a more realistic live potential for the MW. We therefore seek to produce a fiducial model of the Sgr orbit and the associated stream by running $N$-body realizations of the most promising trajectory from the semi-analytic model.  

%%%%%%%%%%%%%%%%%%%%%%%%%%%%%%
\subsection{Parameters and Initial Conditions}
\label{subsec:nbodyparams}

With the simultaneous aims of checking the semi-analytic formalism described earlier and generating a mock Sgr stream, we utilize the $N$-body code {\footnotesize GADGET} \citep{springel05} to re-run the best-fit trajectory presented in Section~\ref{subsec:samresults}. After initially exploiting a simplified spherically symmetric model for the host potential, we now take advantage of having pinned down the best-fit location for the Sun to introduce a flattened disk potential for the MW. With Sgr initialized at (125, 0, 0)~kpc and with $\vec{v}_\text{init} \simeq (-10, 0, 70)$~km~s$^{-1}$, the MW disk inclination is defined by the Sun's position at $\sim(7, -1, 3)$~kpc. The parameters of the Hernquist halo, bulge and exponential disk components of the simulated host and satellite galaxies are summarized in Table~\ref{tab:params}. Since the purpose of this study is to shed light on the dynamical history of Sgr (rather than model its star formation history, for example), the simulation does not include hydrodynamics. The absence of disk gas components should not affect the dynamics of the system, as gas represents only a small percentage of the galaxies' mass budgets. 

Full $N$-body simulations of the MW-Sgr interaction are costly because of the uneven ratio in progenitor mass between the two galaxies. As a result, only a few studies so far \citep[e.g.][]{purcell11, purcell12, gomez15} have modeled the MW with a live halo rather than a static potential. However, given the significant mass of the Sgr satellite and its repeated passages at low Galactocentric distances, it is expected to have strong effects on the structure not only of the MW disk, but also lead to significant dark matter overdensities in the halo \citep{purcell12}. Both the host and satellite haloes are live in our simulation in order to capture such time-dependent effects on the structure of the MW potential. We aim to resolve the total visible mass of Sgr ($\sim 10^9$~M$_{\odot}$ - see Table \ref{tab:params}) to the order of a few tens of thousands of particles in order to sufficiently populate the tidal stream. Therefore, we choose a mass resolution of $4\times10^4$~M$_{\odot}$ per stellar particle, yielding 20,800 stellar particles in Sgr. This in turn implies a required $\sim 2.3\times10^6$ particles to simulate the MW's stellar mass of $\sim 9.4\times10^{10}$~M$_{\odot}$. We use a dark matter particle resolution of $10^6$~M$_{\odot}$, giving $\sim1.2\times10^6$ and $\sim1.2\times10^4$ dark matter particles for the MW and Sgr galaxies respectively. The majority of the computational cost in the simulation therefore comes from adequately resolving the baryonic component of the Sgr dwarf. The option of modeling the MW with fewer particles of higher mass is undesirable, as it leads to rapid two-particle relaxation and introduces artificial disruption of the Sgr satellite \citep{jiang00}. We use an adaptive time step of maximum length 10~Myr, and the softening lengths for the baryonic and dark matter particles are 41~pc and 214~pc, respectively.

%%%%%%%%%%%%%%%%%%%%%%%%%%%%%%
\subsection{Overview of Simulation}
\label{subsec:nbodyoverview}

% Figure: trajectory and separation
\begin{figure*}[hbt]
\begin{center}
\includegraphics[scale=0.5]{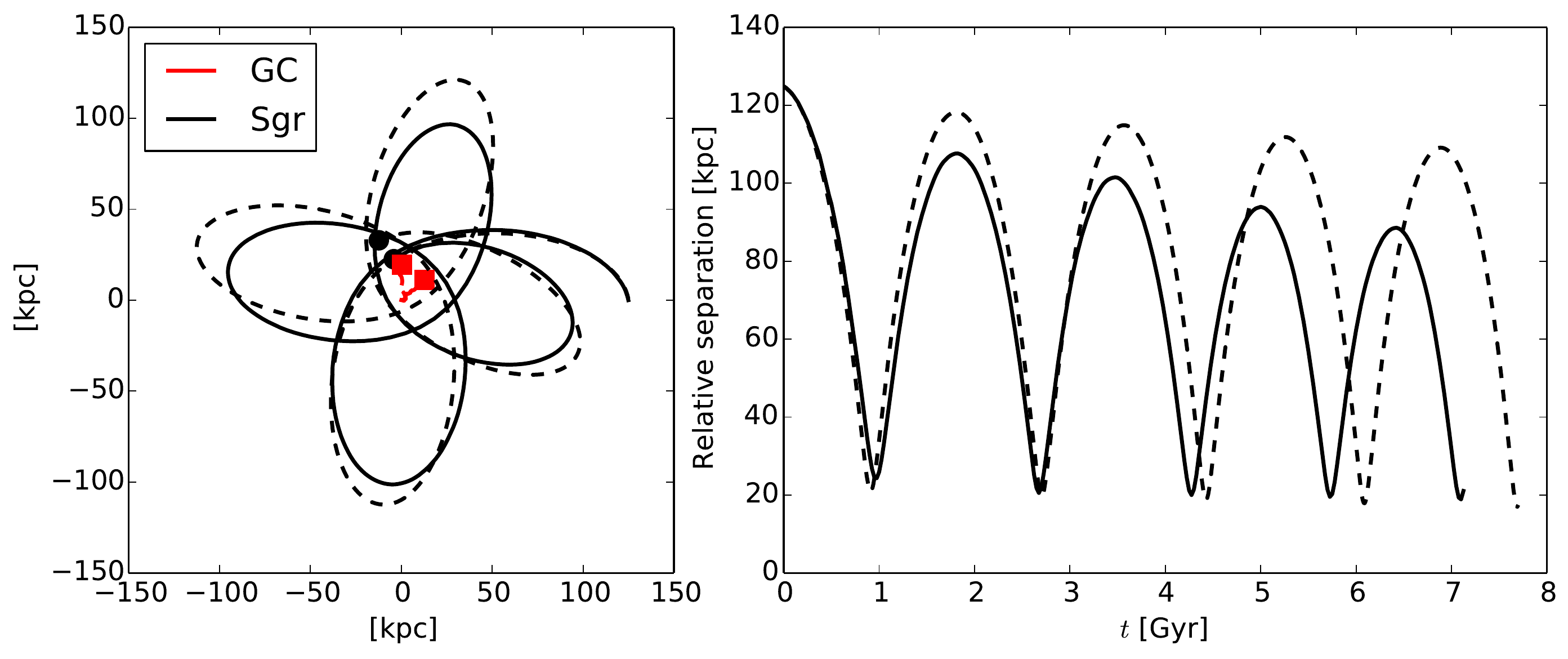} 
\caption{Comparison of the best-fit Sgr orbit computed with {\footnotesize GADGET} and the semi-analytic model described in Section~\ref{sec:SAM} (solid and dashed lines, respectively). {\bf Left panel:} trajectories of the Sgr progenitor (red) and MW Galactic Center (black) in the Sgr orbital plane. The current dynamical centers of the two galaxies are indicated by a colored dot and square, respectively. {\bf Right panel:} separation between the two galaxies as a function of time since the beginning of the calculation. The closest match to the current position and velocity of Sgr is reached after 7.71~Gyr in the semi-analytic model and 7.13~Gyr in the $N$-body run.
\label{fig:trajsep} }
\end{center}
\end{figure*}

% Figure: 3D position and velocity match
\begin{figure*}[hbt]
\begin{center}
\includegraphics[scale=0.5]{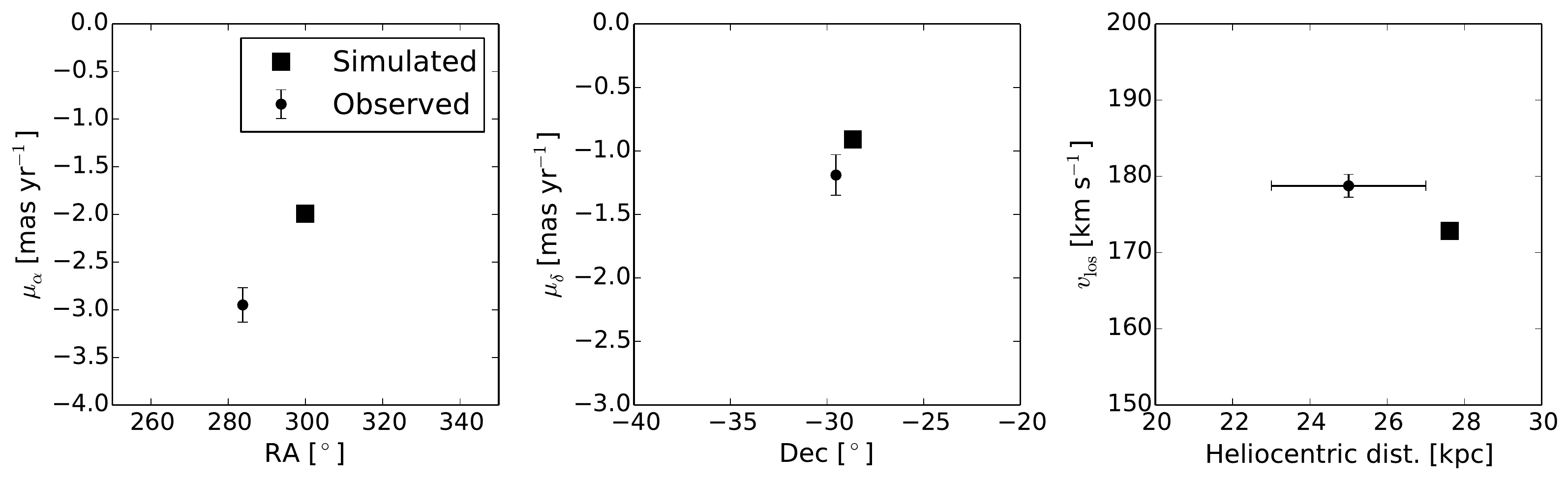} 
\caption{Comparison of the observed coordinates of Sgr to those simulated with an $N$-body code. As in Fig~\ref{fig:SAM3Dmatch}, the position and velocity are shown as projected along the line of sight and in equatorial coordinates on the sky, and error bars show the $1\sigma$ uncertainties associated with each measurement.
\label{fig:3Dmatch} }
\end{center}
\end{figure*}

% Figure: large-scale stream
\begin{figure*}[hbt]
\begin{center}
\includegraphics[scale=0.5]{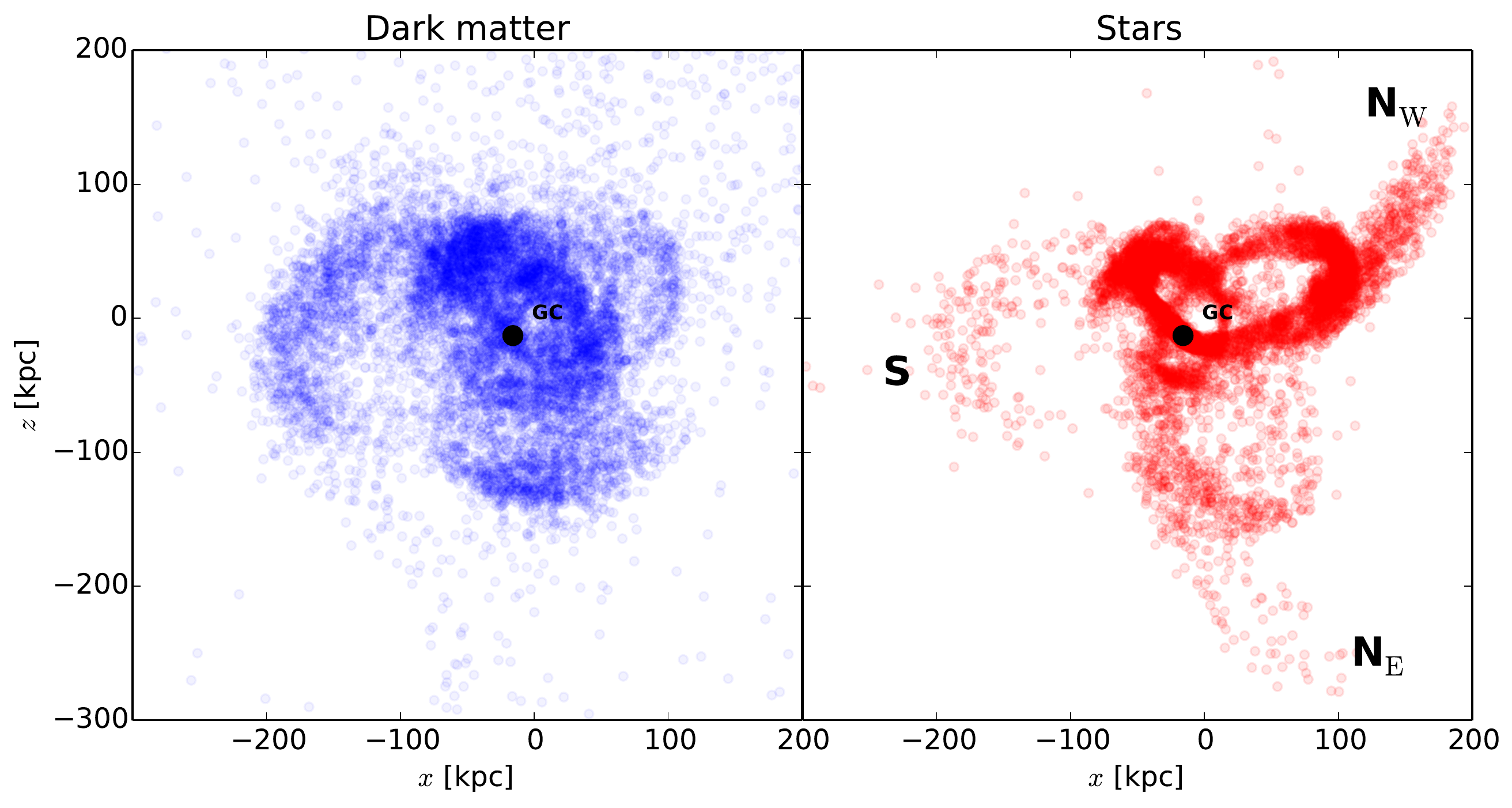} 
\caption{Large-scale view of the Sgr debris stream. The projection plane is defined by its pole at Galactocentric coordinates $(l,b) = (275^\circ, -14^\circ)$, following \citet{belokurov14}. The locations of the Sun and MW center are marked by a black star and square, respectively. {\bf Left panel:} Dark matter particles. {\bf Right panel:} Stellar particles. The three most prominent distant branches of the Sgr stream are labeled N$_\text{W}$ (Northwest), N$_\text{E}$ (Northeast) and S (South).
\label{fig:largestream} }
\end{center}
\end{figure*}

Figure \ref{fig:trajsep} presents a general overview of the orbit of Sgr computed here. The apocenter and pericenter distances decrease at each passage due to dynamical friction. We note that friction is more efficient in the live simulation, gradually shrinking the apocentric distances reached by the satellite as compared to the semi-analytic calculation. Similarly, the discrepancy between the orbital periods in each model grows in time, such that after 7~Gyr of evolution, the timing of the fourth pericenter passage differs by $\sim0.6$~Gyr. In the $N$-body case, the time corresponding to the present-day configuration is reached after 7.13~Gyr. At that time, the match between the observed and simulated phase-space coordinates of Sgr is quantified by a reduced chi-squared value of 2.28, slightly worse than for the semi-analytic model (see Table~\ref{tab:chisquared}). Figure \ref{fig:3Dmatch} shows both the observed and simulated three-dimensional position and velocity of the Sgr core at the present time in relation to the Sun. Across the many test simulations carried out for this work, it is generally the case that the simulated Sgr core is slightly more distant than observed. This challenge in reaching small Galactocentric radii likely arises due to our choice of initial conditions, with the Sgr progenitor starting at larger distances than previously considered for example by \citet{law10} and \citet{purcell11}. Despite the much larger initial separation chosen here, the MW-Sgr distance and relative velocity vectors are still matched within approximately $2\sigma$. Stronger dynamical friction may help reaching smaller Galactocentric separations, suggesting that the Sgr progenitor may have initially been more massive than considered in this study ($M_{\text{Sgr}} = 10^{10}$~M$_\odot$). Testing the dependence of the parameter match on the model MW halo could also provide useful constraints on its properties, a possibility we plan to explore in more depth in a follow-up paper. 

Figure \ref{fig:largestream} presents the behavior of the Sgr progenitor's collisionless components at the time of best match in the simulation. Following \citet{belokurov14}, the coordinates of the stream particles are projected on the debris plane defined by the pole located at $(l_\text{GC}, b_\text{GC}) = (275^\circ, -14^\circ)$. Prominent tidal features are clearly visible for both the dark matter and stellar particles. Shells and arcs corresponding to apocentric pile-ups appear in both the stellar and dark matter component. Interestingly, both the dark and visible particles form large-scale streams at Galactocentric radii well beyond the apocenter distances reached by the Sgr core. The fact that some of the Sgr stars end up outside of the initial orbit can be understood by analogy with the energy redistribution that occurs in tidal disruption events (TDEs). In TDEs, half of the stellar mass gains enough energy to reach the escape speed, while the rest becomes more tightly bound \citep[see e.g.][]{rees88, guillochon16}. The energy gained by the leading arm from the tidal force could lead to large apocentric distances. Our simulation features such extended tidal arms because the initial orbital radius of Sgr is more remote than in previous studies. We have labeled the three most prominent distant extensions N$_\text{W}$, N$_\text{E}$ and S, standing for the northwestern, northeastern and southern branches, respectively. We expect the sharpness of these stellar structures to be affected by the initial distribution of stars in the Sgr model. For a disk-less progenitor, the streams are likely to be less well-defined. Currently, the most distant tidal debris identified are part of the Sgr stream trailing tail, with apocentric distances on the order of $102.5\pm2.5$~kpc according to \citet{belokurov14}. Figure \ref{fig:trajsep} shows that our model predicts significant tidal features at distances $> 100$~kpc. 

%%%%%%%%%%%%%%%%%%%%%%%%%%%%%%
\subsection{Comparison to Stream Data}
\label{subsec:datacomp}

% Figure: Stream overview
\begin{figure*}[hbt]
\begin{center}
\includegraphics[scale=0.5]{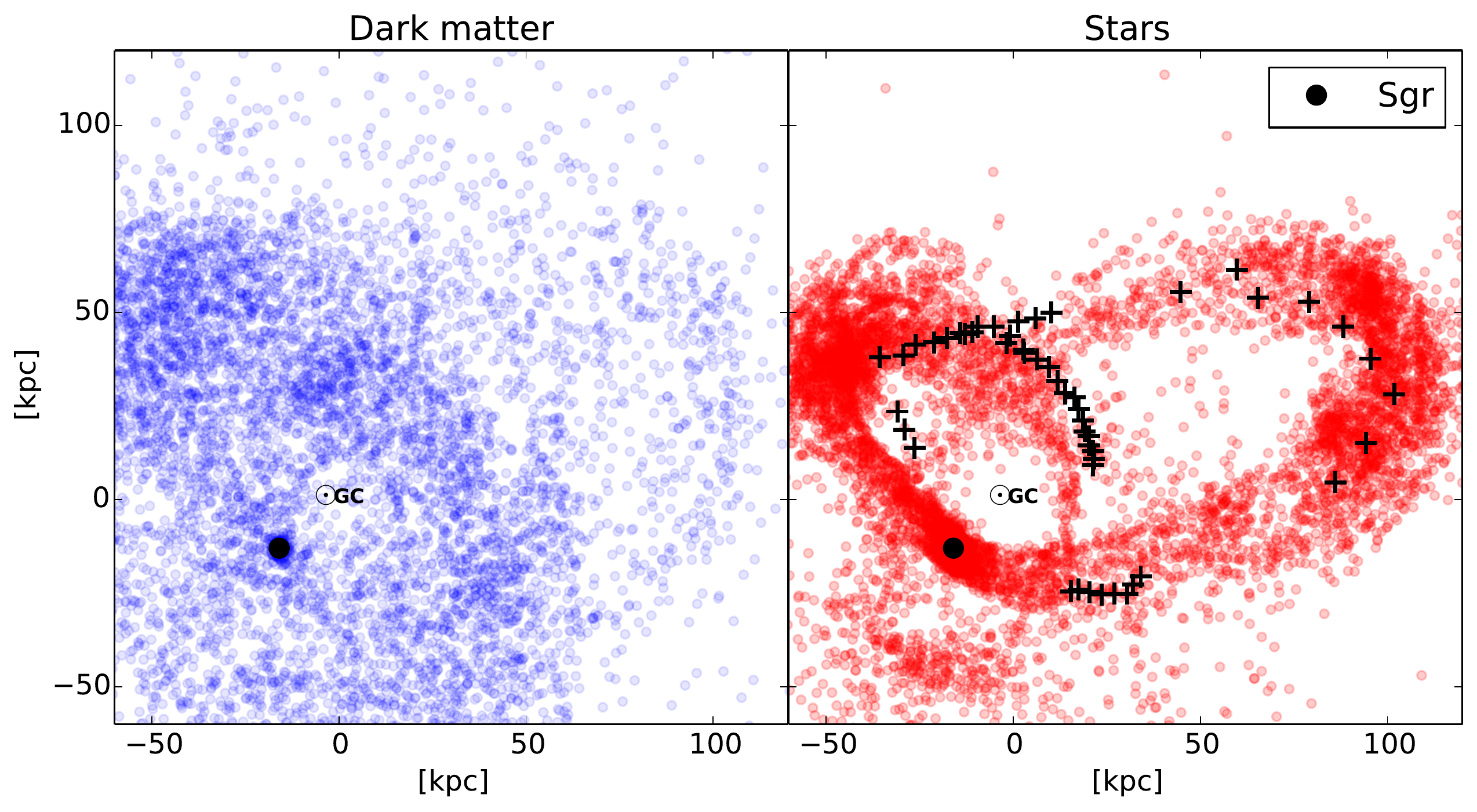} 
\caption{Nearby debris stream projected in the Sgr orbital plane defined by \citet{belokurov14} (as in Fig~\ref{fig:largestream}). The locations of the Sun, MW center and Sgr dwarf remnant are marked by a black star, square and circle, respectively. {\bf Left panel:} Dark matter particles. {\bf Right panel:} Stellar particles. The black plus signs show SDSS data from \citet{belokurov14} Figure 10. The model reproduces both the leading and trailing arms of the stream.
\label{fig:streamoverview} }
\end{center}
\end{figure*}

% Figure: Stream in RA & dec
\begin{figure*}[h!]
\begin{center}
\includegraphics[scale=0.4]{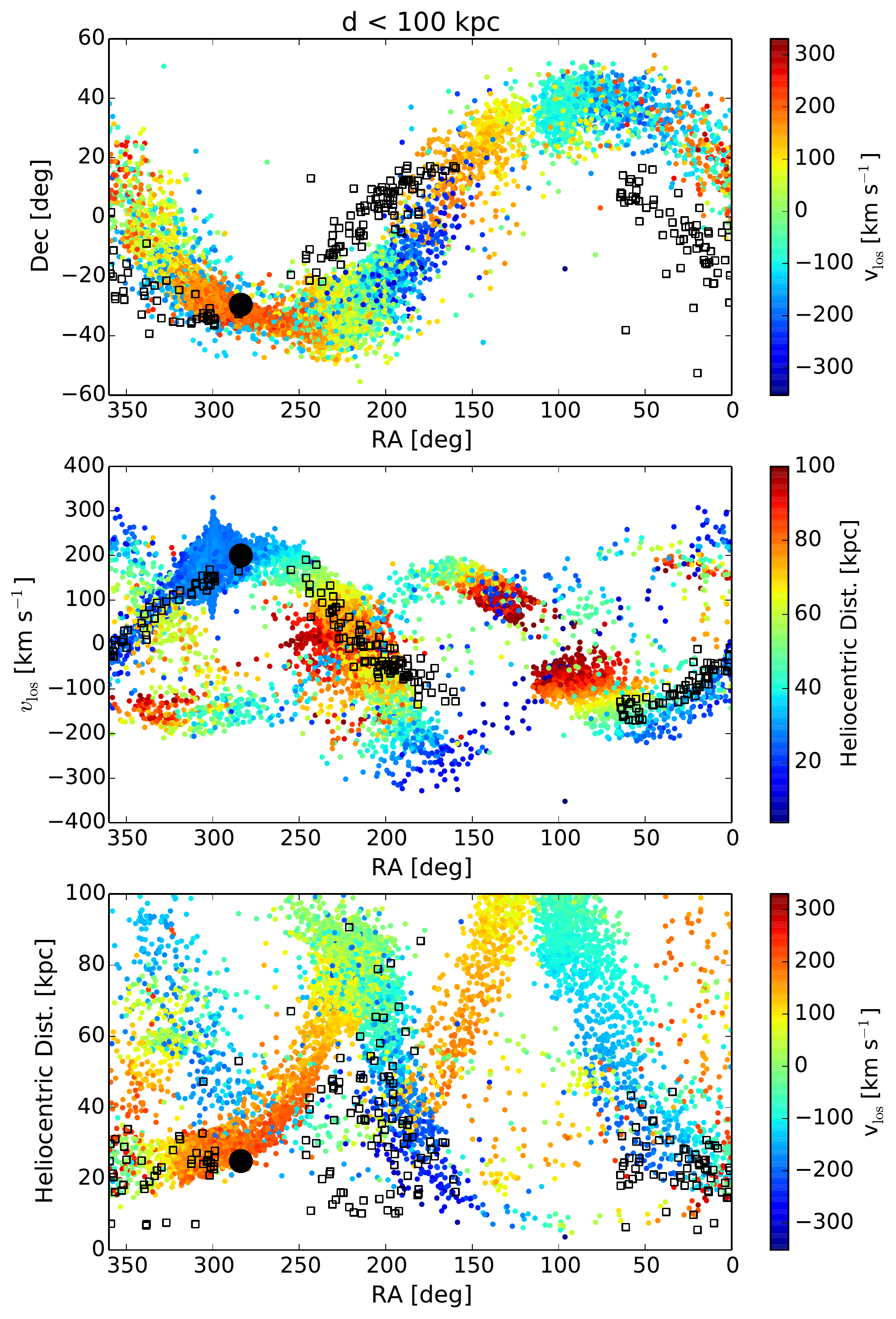} 
\includegraphics[scale=0.4]{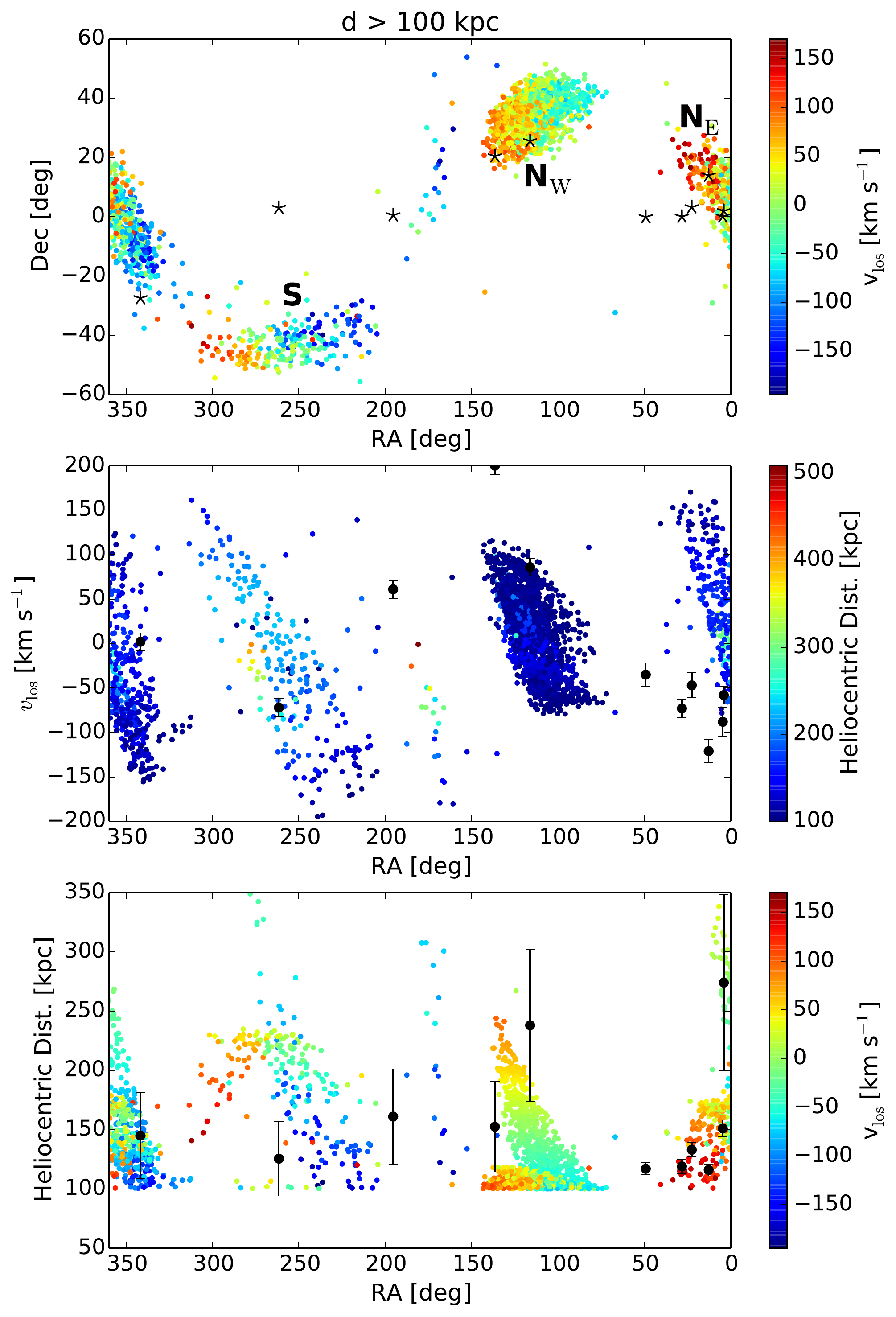} 
\caption{Equatorial coordinates, heliocentric distance and Galactic Standard of Rest line-of-sight velocity of the Sgr stellar particles at the present time. The simulated stars are color coded according to distance or line-of-sight velocity, as indicated in the color bars. The observed location of the Sgr remnant core is marked by a black circle. {\bf Left panels:} Stellar particles with distances $<$100 kpc. Black squares show data for M-giant stars from the Two Micron All Sky Survey (2MASS) from \citet{majewski04}. We have chosen axis ranges identical to those in \citet{gomez15} (Figure 8) to allow direct comparison with the N-body model used in their study, which is based on the work of \citet{purcell11}. {\bf Right panels:} Stellar particles with distances $>$100~kpc. The features located at RA~$\sim 100^\circ$, $\sim 5^\circ$, and $\sim 250^\circ$ correspond to the distant N$_\text{W}$, N$_\text{E}$, and S branches, respectively.  The distant stars detected by \citet{deason12a} and \citet{bochanski14b} are displayed as black markers. Note the smaller range of line-of-sight velocities for these distant stars, captured closer to orbital turnaround. Of the 11 distant detections plotted here, approximately half coincide with the predicted Sgr stream structure.
\label{fig:RAdec} }
\end{center}
\end{figure*}

In Figure \ref{fig:streamoverview} we present a comparison between available data and the modeled stream features projected on the Sgr orbital plane as defined by \citet{belokurov14}. In the right panel, black crosshairs show data from Sloan Digital Sky Survey (SDSS) Data Release 5 and 8 as aggregated in Figure 10 of \citet{belokurov14}. The size of the markers are indicative of the largest distance errors present in the dataset. Our choice of initial conditions for the Sgr baryonic components likely affects the thickness of the simulated stream. We did not try to vary the velocity dispersion of the stellar particles to optimize for stream width.

Both the leading and trailing arm SDSS detections have obvious counterparts in the simulation. The leading branch of the simulated stream presents a small offset both with the observed data and the model by \citet{law10} shown in \citet[][Fig. 10]{belokurov14}: the apocenter distance appears $\sim10$~kpc larger than the measured value of $47.8 \pm 0.5$~kpc \citep{belokurov14}, representing a $\sim20$\% mismatch. The position angle of the apocenter location is also slightly different from the data. \citet{belokurov14} have argued that the precession angle of only $93.2^\circ \pm 3.5^\circ$ measured from the SDSS data is an indication that the MW dark matter density falls more quickly with radius than a logarithmic potential (with typical precession angles of $120^\circ$). While orbital energy and angular momentum also play a minor role, the fact that the simulated precession angle appears $\sim10^\circ$ larger than the measured value might suggest a steeper model potential is needed for the MW. 

However, the model's most promising feature is that it successfully reproduces the distant trailing arm of the Sgr stream. Initially detected by \citet{newberg03}, remote Sgr stellar debris in the Northern hemisphere were later confirmed by \citet{ruhland11}, \citet{drake13}, \citet{belokurov14}, \citet{koposov15}, and \citet{li16}. Earlier studies hesitated to assign this structure to the Sgr stream, as simulations did not show any counterparts to these distant stars \citep[e.g.][]{ruhland11, drake13}. Importantly, our model correctly replicates both the measured apocenter distance of $102.5 \pm 2.5$~kpc and the position angle found by \citet{belokurov14}. Two main mechanisms have been invoked in the literature to explain the difference in apocentric distances of $\sim50$~kpc measured by \citet{belokurov14}. According to \citet{chakrabarti14}, the disparity is due to dynamical friction modulating the eccentricity of the orbit of Sgr. Given the high initial separation/progenitor mass regime investigated in our simulation, dynamical friction plays an important role in shrinking down the pericenter and apocenter distances reached by the Sgr dwarf. On the other hand, \citet{gibbons14} find that the stream behavior depends primarily on the host potential and secondarily on the satellite progenitor properties. 

Figure \ref{fig:RAdec} shows the present-day distribution of the simulated Sgr stars in different projections of phase-space. The left-hand-side panels feature all Sgr star particles inside a Galactocentric radius of 100~kpc, along with stellar tracer data from the Two Micron All Sky Survey (2MASS) from \citet{majewski04}. Overall, the simulation produces a reasonable match to the observed tidal debris characteristics shown in black squares. Unlike the stellar distribution presented in \citet{gomez15}, we successfully reproduce the distant branch of the leading arm observed at RA~$\sim 200^\circ$ and distances $>50$~kpc. The model however predicts comparatively fewer stars at smaller distances. This comparison is limited, however, by the low number of Sgr stellar particles in the simulation and the resulting relatively sparse sampling. Similarly, the simulated data at RA~$\sim 150^\circ$ and dec~$\sim 20^\circ$ are too coarse to distinguish the presence of the bifurcation in the leading stream detected by \citet{belokurov06}. Observational and theoretical studies \citep[e.g.][]{belokurov14, penarrubia10, penarrubia11} have not yet reached a consensus regarding the dependence of the bifurcated tails on the morphology of the MW and the Sgr progenitor. More comprehensive simulations and detailed testing of galactic model parameters are needed in order to shed light on the origins of the stream bifurcation.

The simulated stellar locations in (RA, dec) feature a systematic offset compared to the data, and the location of highest particle density does not line up exactly with the Sgr centroid. Such coherent shifts of the simulated stream occasionally appeared across the numerous test simulations carried out for this study. The combination of coarse sampling of possible Sun locations on the Fibonacci sphere (see Section \ref{subsec:sammethods}) as well as the time interval of 25~Myr between subsequent output snapshots may be responsible for these offsets. Finer sampling and tuning of the simulation parameters could perhaps remove this issue. Alternatively, this angular offset may indicate that the simple spherical potential used to model the MW halo in this simulation is insufficient.

Additionally, the run of line of sight velocities for the leading arm ($250^\circ \lesssim \text{RA} \lesssim 150^\circ$) features a discrepancy with the data similar to that discussed in other studies \citep[e.g.][]{helmi04, law05}. \citet{helmi04} suggest that a prolate model of the MW potential is required to solve this mismatch. However, \citet{law05} and \citet{johnston05} point out that such models introduce a discordance with the precession angles measured from stream M-giants. While these studies disagree on the exact nature of the MW triaxiality (oblate vs. prolate), there is a consensus that it is difficult to reproduce the leading arm radial velocities with a simple spherically symmetric model (such as the one used in our study). On the other hand, some papers \citep{law10, gomez15} have shown that the LMC can introduce significant perturbations to the phase-space distribution of Sgr debris. These hypotheses could be tested in future iterations of the orbital model presented in our study. 

The set of panels on the right of Figure~\ref{fig:RAdec} shows the predicted distribution of stars beyond distances of 100~kpc, where no Sgr stream debris have yet been identified with certainty. The large-scale tidal features labeled in Figure~\ref{fig:largestream} appear as prominent clouds of stars in these panels. The structures located at RA~$\sim 100^\circ$, $\sim 5^\circ$, and $\sim 250^\circ$ correspond to those labeled N$_\text{W}$, N$_\text{E}$, and S in Figure~\ref{fig:largestream}, respectively. These distant branches can be distinguished from the closer debris in that area of the sky through line-of-sight velocities (color-coded with a different range for the distant stars), which are generally lower by $\sim50$~km~s$^{-1}$ since they approach turnaround. 

While no tidal streams were so far identified at such large distances \citep{drake13}, Figure~\ref{fig:RAdec} includes the parameters of the few MW stars found at distances $>100$~kpc. A variety of stellar tracers have been used to map the inner halo. However, due to faint limits on the order of $r\lesssim21$, surveys so far have only yielded a dozen or so stars beyond 100~kpc. \citet{deason12a} provide a sample of distant blue horizontal branch and N-type carbon stars, later complemented by the M-giants detected by \citet{bochanski14b}. In their UKIRT Infrared Deep Sky Survey (UKIDSS) data, \citet{bochanski14b} were able to identify two stars with distances above 200~kpc, making them the most distant known MW stars. Figure~\ref{fig:RAdec} shows that these most distant stars appear consistent with the remote northern stream branch predicted in our simulation. \citet{bochanski14b} already associated the M-giants in a distance range of 20-90~kpc in the UKIDSS data with Sgr. However, the more distant detections in their sample were not connected with the stream because the models of \citet{law10} do not feature stars beyond $\sim75$~kpc. Our study shows that some of the most distant known MW stars could have originated in Sgr.

%%%%%%%%%%%%%%%%%%%%%%%%%%%%%%
\section{Discussion and Conclusions}
\label{sec:conclusions}

We have taken a two-pronged approach to building a new model of the Sgr orbit through the MW halo. Rather than integrate the present-day phase-space coordinates backwards in time, we perform an exploration of parameter space with a semi-analytic integration of the Sgr trajectory starting 7-8~Gyr ago. This method is chosen because it allows for the inclusion of non time-reversible effects. In particular, dynamical friction and tidal stripping are expected to be important in the regime where the mass of the Sgr progenitor is $\gtrsim10^{10}$~M$_\odot$. We then build on these results to simulate the trajectory of Sgr with {\footnotesize GADGET} and compare the resulting tidal stream to observational data. Our main conclusions are as follows:

\begin{enumerate}
\item Comparing the simulated position and velocity of Sgr today to the measured quantities, we find that our analytic model favors a narrow range of the initial orbital angular momentum of Sgr, with large incidence angles ($\theta_\text{init}\gtrsim 60^\circ$) and initial velocities in the range $80$~km~s$^{-1} \lesssim v_\text{init} \lesssim 100$~km~s$^{-1}$. 
\item The mock Sgr stream resulting from the {\footnotesize GADGET} simulation reproduces most of the 2MASS \citep{majewski04} sky positions, heliocentric distances, and line of sight velocities. Similarly to previous studies, the leading arm line of sight velocities are not replicated by the model, suggesting that further work on the triaxiality of the MW halo or the inclusion of the LMC influence is necessary.
\item The simulated debris stream projected on the plane of the Sgr orbit is in unprecedentedly good agreement with the SDSS stellar tracers from \citet{belokurov14}. The stream apocentric distances and position angles are reproduced to within 20\% of the measured values. In particular, the model of the Sgr orbit presented here is the first to naturally reproduce the recently detected distant trailing arm at $\sim100$~kpc, optimizing only for the remnant coordinates rather than fitting for the stream properties. We believe that this feature arises because of the larger initial separation (125~kpc) used in our simulation compared to other works in the literature \citep[e.g.][]{law10, purcell11, gomez15}.
\item Above all, this work predicts the existence of two novel and distant arms of the Sgr stream. Currently the most distant SDSS stream detections are located at distances of $\sim102.5\pm2.5$~kpc \citep{belokurov14}. The simulation presented here includes stellar overdensities at distances of up to $\sim250-300$~kpc, extending beyond the MW virial radius. We provide their predicted positions on the sky, heliocentric distances and line of sight velocities for possible future observational searches. The most distant known stars of the MW coincide with our predicted streams in both position and small radial velocities. If verified observationally, the distant branches of the Sgr stream would be the farthest-ranging stellar stream in the MW halo known to date.
\end{enumerate}
UKIDSS is sensitive to M-giants beyond the MW virial radius and its distant detections appear consistent with our predicted stream. Further findings within existing datasets may be possible with the Panoramic Survey Telescope \& Rapid Response System\footnote{http://pan-starrs.ifa.hawaii.edu/public/} (Pan-STARRS) forced photometry method, or with the Dark Energy Camera Legacy Survey\footnote{http://legacysurvey.org/decamls/} (DECaLS). In the future, the high depth of Large Synoptic Survey Telescope\footnote{https://www.lsst.org} (LSST) data will allow detailed mapping of the outer halo in the visible band, while the Wide Field Infrared Survey Telescope\footnote{http://wfirst.gsfc.nasa.gov} (WFIRST) will improve on current UKIDSS photometry in the infrared. The detection and characterization of the distant branches of the Sgr stream would provide an unprecedented opportunity to probe the outer envelope of the MW, which is also influenced by the neighboring Andromeda galaxy. The {\it Gaia} mission will accurately map a large volume of the MW at smaller Galactocentric radii. With a complete picture of the MW mass distribution from the solar neighborhood to the outskirts of the halo, we will be able to place our Galaxy and the Local Group in a cosmological context. 

\acknowledgments
We thank Laura Blecha, Vasily Belokurov, Gurtina Besla and Facundo G\`{o}mez for valuable discussions, as well as the referee for helpful comments on the manuscript.

\end{document}